\documentclass[useAMS,usenatbib]{mn2e}

\usepackage{graphicx}
\usepackage{bm}

\title[New group of double-periodic RR Lyrae stars]{Discovery of a new group of double-periodic RR Lyrae stars in the OGLE-IV photometry}
\author[H. Netzel, R. Smolec \& W. Dziembowski]
{H. Netzel$^{1}$\thanks{E-mail: henia@netzel.pl},
R. Smolec$^{2}$\thanks{E-mail: smolec@camk.edu.pl} and
W. Dziembowski$^{2,3}$\\
$^{1}$Instytut Astronomiczny, Uniwersytet Wroc\l{}awski, ul. Kopernika 11, 51-622 Wroc\l{}aw, Poland\\
$^{2}$Nicolaus Copernicus Astronomical Centre, Polish Academy of Sciences, Bartycka 18, 00-716 Warszawa, Poland\\
$^{3}$Warsaw University Observatory, Al. Ujazdowskie 4, 00-478 Warszawa, Poland\\
}

\begin{document}

\date{Accepted . Received ; in original form }

\pagerange{\pageref{firstpage}--\pageref{lastpage}} \pubyear{2014}

\maketitle

\label{firstpage}

\begin{abstract}
We report the discovery of a new group of double-periodic RR Lyrae stars from the analysis of the OGLE-IV Galactic bulge photometry. In 11 stars identified in the OGLE catalog as first overtone pulsators (RRc stars) we detect additional longer period variability of low amplitude, in the mmag regime. One additional star of the same type is identified in a published analysis of the {\it Kepler} space photometry. The period ratio between the shorter first overtone period and a new, longer period lies in a narrow range around $0.686$. Thus, the additional period is longer than the expected period of the undetected radial fundamental mode. The obvious conclusion that addition periodicity corresponds to a gravity or a mixed mode faces difficulties, however. 

\end{abstract}


\section{Introduction}\label{sec:intro}
\begin{figure}
\centering
\resizebox{\hsize}{!}{\includegraphics{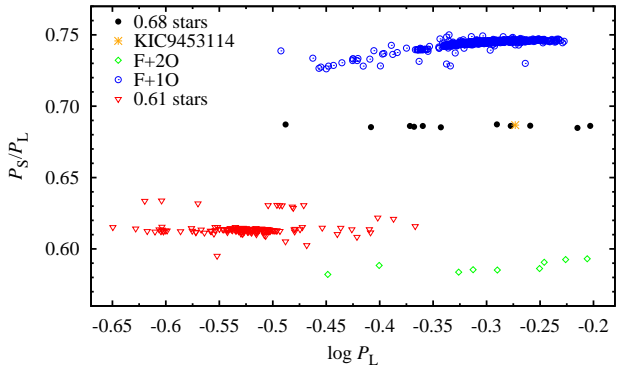}}
\caption{Petersen diagram for multiperiodic RR~Lyrae stars. $P_{\rm L}$ and $P_{\rm S}$ are longer and shorter period, respectively, and correspond to different pulsation modes in the groups plotted in the diagram.}
\label{fig.pet}
\end{figure}

RR~Lyrae stars are classical pulsators of great astrophysical importance. They serve as excellent distance indicators and tracers of the old stellar populations, enabling galactic structure, kinematics and evolution studies. Majority of these stars are radial, single-periodic pulsators, pulsating either in the fundamental mode (F-mode, RRab stars) or in the first overtone (1O-mode, RRc stars). Double-mode pulsators pulsating simultaneously in the fundamental and in the first overtone modes (RRd stars) are also known and numerous. In the Petersen diagram, i.e. a diagram of shorter-to-longer period ratio vs. the longer period these stars form a well defined group with characteristic period ratio in a range $0.725-0.747$ \citep[depending on the period; open circles in Fig.~\ref{fig.pet},][]{ogleiv}. In all groups the Blazhko effect, quasi-periodic modulation of pulsation amplitude and phase is observed \citep[for a review see][]{szabo14}. The effect is more frequent in RRab stars than in RRc stars. Only recently the effect was discovered in RRd stars \citep{ogleiv,jurcsik_BLRRd, rs15a}. RRc stars are prone to fast and irregular period changes on a time-scales shorter than predicted from stellar evolution theory \citep[e.g.][]{csbook}.

Recently, new groups of double-periodic RR~Lyrae stars were identified, thanks to precise and nearly continuous photometry of the space missions, {\it CoRoT} and {\it Kepler}. The discoveries include fundamental plus second overtone radial mode pulsators [e.g. \cite{benko10}, for a review see \cite{pam14}; diamonds in Fig.~\ref{fig.pet}] and mysterious group of double-mode radial-non-radial pulsators. In this group we observe the dominant pulsation in the first overtone mode and additional variability with a shorter period. In the Petersen diagram (triangles in Fig.~\ref{fig.pet}), these stars seem to form two groups with period ratios clustering around $\approx0.61$ and $\approx0.63$. Space photometry \citep[][]{szabo_corot,pamsm15,molnar} indicates that this form of pulsation must be common among RRc stars, as 13 out of 14 RRc stars observed from space show the phenomenon. Majority of the 0.61 stars plotted in Fig.~\ref{fig.pet} however, were detected only recently in the third phase of the Optical Gravitational Lensing Experiment \citep[OGLE; see eg.][]{ogleIII} photometry of the Galactic bulge by \cite{netzel}. 
We also note that in RRc stars observed from space other low frequency signals were detected and interpreted as non-radial gravity modes \citep{pamsm15}.

In this Letter we describe the discovery of a new group of double-periodic pulsators among stars identified as RRc. In the analysis of a top-quality sample of OGLE-IV photometry of the Galactic bulge \citep{ogleiv} we have found 11 stars that show yet another period ratio that cannot be explained with simultaneous excitation of two radial modes. These stars are marked with filled circles in Fig.~\ref{fig.pet}. The additional periodicity is longer than the first overtone period and, as comparison with RRd stars clearly reveals, much longer than the expected period of the undetected radial fundamental mode. A literature search revealed yet another member of the group in the {\it Kepler} photometry of RRc stars \citep{pamsm15}. In the following sections we describe our analysis and summarize the properties of the new group. In Section~\ref{sec:summary} we briefly consider possible explanation for the group.

\section{Data and analysis}\label{sec:analysis}

Data used for analysis were collected during the fourth phase of the OGLE project. For some stars we used additionally data from the third phase. OGLE-IV collection of variable stars \citep{ogleiv} contains $38\,257$ RR~Lyrae stars, including $10\,825$ RRc stars and $174$ RRd stars. Our goal is to search for additional, low amplitude signals in stars pulsating in the first overtone (RRc and RRd). Detecting weak signals requires possibly the lowest noise level, which is best met by stars with the largest number of observations. Hence, for our analysis we chose stars located in the most frequently observed fields, their numbers are 501 and 505 in OGLE-IV. For each star in these fields we have more than $8\,000$ datapoints. Altogether, there are 485 RRc and 4 RRd stars in these fields, which is also a reasonable number for manual analysis. We used observations in the $I$-band only (more numerous than the $V$-band data). In publicly available OGLE-IV data there are four observational seasons available. Length of the data is about $1334$\thinspace d for most stars.

Data were analysed manually using standard consecutive prewhitening method. Discrete Fourier transform allows to find significant frequencies in the spectrum. We considered only those frequencies for which signal-to-noise ratio was $S/N \geqslant 4$. All detected frequencies were fitted to the data in the form:
\begin{equation} \label{eq.fit}
 m(t)=m_0+\sum_{k=1}^N{A_k\sin(2\pi \nu_k t + \phi_k)}\,,
\end{equation}
where $\nu_k$ are frequencies, $A_k$ and $\phi_k$ are amplitudes and phases. Only resolved frequencies are included in (\ref{eq.fit}). We consider two frequencies unresolved if separation between them $\Delta\nu<2/T$, where $T$ is length of the data. When all significant frequencies were included in the sum, we removed points deviating from the fit by more than $4\sigma$.

In the data of many stars slow trend is present. It manifests in the frequency spectrum as a signal at low frequencies and gives rise to daily aliases at integer frequency values. We model the slow trends either with a long-period ($\approx 50\,000$\thinspace d) sine function or with low-order polynomial.

In many stars, after prewhitening with the frequency of the first overtone, $\nu_{\rm 1O}$, and its harmonics, close peaks were detected at $k\nu_{\rm 1O}$, forming either equidistant triplets (multiplets) or close doublets. These are a signature of the Blazhko effect. We fitted these signals in the form $k\nu_{\rm 1O} \pm \Delta\nu$, where $\Delta\nu$ is a separation between main frequency and the side peaks.

In other stars, after prewhitening, unresolved signal remains at the location of $\nu_{\rm 1O}$ and/or its harmonics. It indicates that first overtone is not stationary, but its amplitude and/or phase change with time on a time scale comparable to or longer than the data length.  Such signals increase the noise level in the transform and may hide the additional low amplitude signals, which we search for. To get rid of the non-stationary signals we used time-dependent prewhitening method proposed by \cite{pamsm15} \citep[see][for application to OGLE data]{netzel}. Whenever possible, in order to investigate long-term variation of the first overtone, we combined OGLE-IV data with OGLE-III data \citep{ogle_rr_blg}, which, in some cases revealed a long-period Blazhko modulation. Irregular phase (period) changes of the first overtone are also very frequent.

\section{Results}\label{sec:results}
As a result of our study we identified several interesting and well known phenomena: Blazhko effect, period-changing stars and double-periodic stars with additional non-radial, shorter period mode, with characteristic $\approx 0.61$ period ratio to the first overtone period. These results will be described elsewhere (Netzel et al., in prep.). In addition, we found a group of 11 stars ($2\%$ of the sample) with additional low frequency signal. Corresponding period is longer than first overtone period; period ratios, $P_{\rm 1O}/P_{\rm x}$ fall within a narrow range from $0.6848$ to $0.6872$ with average value of $0.6860$. Such period ratios were not reported in the literature before. Properties of the stars are summarized in Tab.~\ref{tab:sh}.

 A literature search revealed one additional RRc star observed by {\it Kepler} in which additional longer period was detected with period ratio falling in the same range \citep[see $f_5$ in tab.~7 in][]{pamsm15}. Data for the star are in the last row of Tab.~\ref{tab:sh}.

\begin{figure}
\centering
\resizebox{\hsize}{!}{\includegraphics{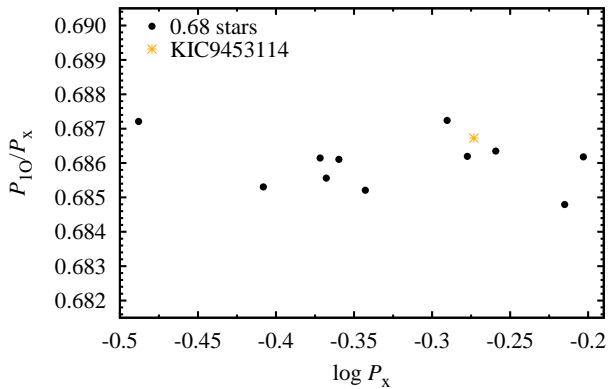}}
\caption{Petersen diagram for stars with additional frequency.}
\label{fig.pet068}
\end{figure}

Stars with the additional frequency are plotted with filled circles  on the Petersen diagram in Fig.~\ref{fig.pet}. Star found in the {\it Kepler} photometry is marked with a different symbol and it fits the progression formed by the OGLE stars very well. All stars form a tight horizontal sequence below RRd stars. Fig.~\ref{fig.pet068} shows only the newly found stars. No clear structure is visible within the group.

 We note that period ratios of RRd stars cover a wide range in the Petersen diagram with period ratio clearly correlated with the period of the fundamental mode. The large range of period ratios of RRd stars corresponds to the large metallicity spread in the Galactic bulge, as model calculations indicate \citep[eg.][]{ogle_rr_blg}. Without identification of the additional pulsation mode and appropriate pulsation models (see Sec.~\ref{sec:summary}) we cannot infer about metallicities or relation between metallicity and location of the star in the Petersen diagram for the newly discovered group.

The analysed stars are located in the most dense stellar fields in the Galactic bulge, in which probability of blending is very high. The tight progression formed by a new group in the Petersen diagram and nearly constant period ratio with the first overtone period are a strong argument that the additional signal is real and intrinsic to the analysed RRc stars. Even more, in six stars we detect combination frequencies of the additional and first overtone frequencies which proves that the two signals originate from the same star. The form of the combination frequency is written in the penultimate column of Tab.~\ref{tab:sh}. In three stars $S/N>4$ for signal at combination frequency (marked with bold font). In other three stars we see a signal precisely at the position of combination frequency (within frequency resolution of the spectrum), but with $3<S/N<4$.

\begin{figure}
 \centering
\resizebox{\hsize}{!}{\includegraphics{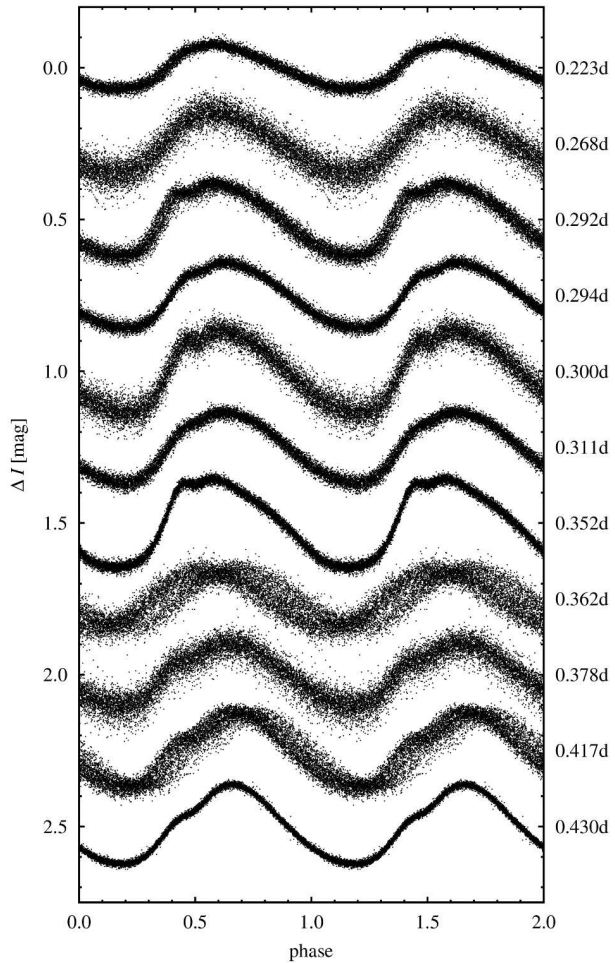}}
  \caption{Light curves of stars with additional frequency phased with a period of the first overtone. They are ordered according to increasing period.}
    \label{fig.light_curves}
 \end{figure}

The unknown period is longer than the period of the unseen fundamental mode (see Fig.~\ref{fig.pet}). This has an immediate consequence: if additional variability corresponds to pulsation mode, it can not be radial. Even more, it can not correspond to the purely acoustic mode, but must be a gravity mode or a mode of mixed character. As such explanation faces difficulties (Sec.~\ref{sec:summary}) we analysed the light curves of the dominant pulsation mode to check whether the RRc identification in the OGLE catalog is correct. The light curves folded with the first overtone period are plotted in Fig.~\ref{fig.light_curves} sorted by the increasing period. Shapes are typical for RRc stars including the characteristic bump-feature preceding the maximum light. Period change, well visible for some longer period stars, is also characteristic for RRc stars. It also contributes to the larger photometric dispersion of some of the light curves. The other factor responsible for different dispersion of the light curves presented in Fig.~\ref{fig.light_curves} is difference in mean brightness of the stars: dispersion is larger for fainter stars.

\begin{figure}
 \centering
\resizebox{\hsize}{!}{\includegraphics{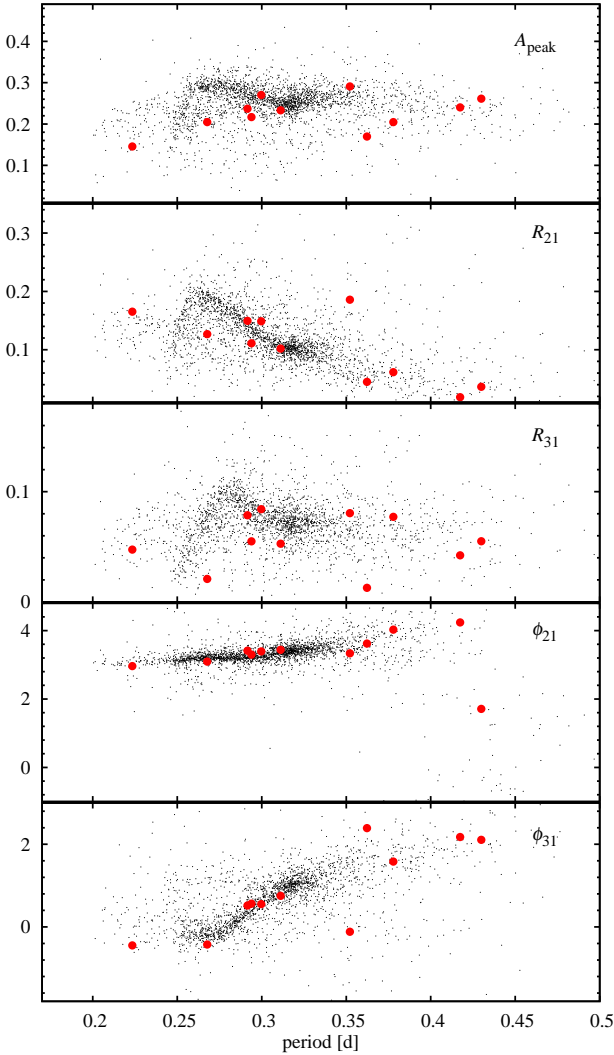}}
  \caption{Fourier decomposition parameters for RRc Galactic bulge stars (OGLE-III). Red points correspond to stars with the additional frequency. Rest of the RRc stars are marked with black dots.}
   \label{fig.fco}
 \end{figure}

Light curve shapes can be described quantitatively with the Fourier decomposition parameters, which display characteristic progression with the pulsation period, depending on the pulsation mode. In Fig.~\ref{fig.fco} we compare the Fourier decomposition parameters for our stars with parameters for RRc stars from the Galactic bulge and OGLE-III catalog \citep{ogle_rr_blg}. Errors in determination of Fourier decomposition parameters for our stars are usually smaller than symbol size. Parameters for all 11 stars are within typical for RRc stars at given pulsation period. Based on the photometric data we have, we conclude that the dominant pulsation mode is indeed the radial first overtone.

In 9 stars we see a signal close to the primary frequency. Those stars are marked with `a' in the remarks column, regardless of whether the signal is resolved or not. These stars were investigated for the Blazhko effect using combined OGLE-III and OGLE-IV data (see next Section). Typically irregular phase changes were detected. In one star, marked with `b', we found harmonic of the additional mode. In one star, marked with `c', signal close to additional frequency is seen (see next Section). In all stars additional signal is stationary.

\begin{table*} 
 \centering 
 \caption{Stars with additional frequency. Subsequent columns contain periods of the first overtone and of additional signal, their ratio, amplitude of the first overtone and amplitude ratio. Two last columns contain form of the combination frequencies (if detected) and remarks.} 
 \label{tab:sh} 
 \begin{tabular}{lr@{.}lr@{.}lrrrrr} 
 \hline 
 star & \multicolumn{2}{c}{$P_{\rm 1O}$ (d)} & \multicolumn{2}{c}{$P_{\rm x}$ (d)} &$P_{\rm 1O}/P_{\rm x}$& $A_{\rm 1O}$ (mag) & $A_{\rm x}/A_{\rm 1O}$ & combination freq. & remarks\\ 
 \hline
OGLE-BLG-RRLYR-04994 & 0&3622954(3) & 0&52797(1) & 0.68619 & 0.0799(4) & 0.048 &  & a \\
OGLE-BLG-RRLYR-05080 & 0&2996982(1) & 0&436810(5) & 0.68611 & 0.1353(4) & 0.048 & $\bm{\nu_{\rm 1O}+\nu_{\rm x}}$ &  \\
OGLE-BLG-RRLYR-06970 & 0&42988998(7) & 0&626497(9) & 0.68618 & 0.1272(1) & 0.012 & $\nu_{\rm 1O}+\nu_{\rm x}$ & a \\
OGLE-BLG-RRLYR-07127 & 0&3778854(3) & 0&55057(1) & 0.68635 & 0.1009(5) & 0.050 &  & a \\
OGLE-BLG-RRLYR-07653 & 0&31118879(6) & 0&454151(6) & 0.68521 & 0.1166(2) & 0.025 &  & a \\
OGLE-BLG-RRLYR-08748 & 0&29153824(7) & 0&424892(3) & 0.68615 & 0.1185(3) & 0.039 &  & a,b \\
OGLE-BLG-RRLYR-09146 & 0&35215799(3) & 0&5124233(5) & 0.68724 & 0.1474(1) & 0.028 & $\bm{\nu_{\rm 1O}+\nu_{\rm x},2\nu_{\rm 1O}+\nu_{\rm x}}$ & c \\
OGLE-BLG-RRLYR-09217 & 0&29391346(2) & 0&428719(1) & 0.68556 & 0.1083(2) & 0.026 &  $\nu_{\rm 1O}+\nu_{\rm x}$ & a \\
OGLE-BLG-RRLYR-09426 & 0&22339120(4) & 0&325069(6) & 0.68721 & 0.0724(2) & 0.016 & $\nu_{\rm 1O}+\nu_{\rm x}$ & a \\
OGLE-BLG-RRLYR-10100 & 0&4173796(2) & 0&609497(8) & 0.68479 & 0.1164(4) & 0.060 & $\bm{\nu_{\rm 1O}+\nu_{\rm x}}$ & a \\
OGLE-BLG-RRLYR-32196 & 0&2677486(1) & 0&390699(7) & 0.68531 & 0.1008(4) & 0.039 &  & a \\
\hline 
 KIC9453114 & 0&3660809 & 0&5330831 & 0.68672 & 0.20664 & 0.004 & $\nu_{\rm x}-\nu_{\rm 1O}$, $\nu_{\rm x}-2\nu_{\rm 1O}$ & \\ 
 \hline 
 \multicolumn{7}{l}{a -- additional signal close to $\nu_{\rm 1O}$; b -- harmonic of $\nu_{\rm x}$; c -- additional signal close to $\nu_{\rm x}$}\\ 
 \hline 
 \end{tabular} 
 \end{table*}

\section{Remarks on individual stars}\label{sec:remarks}
{\bf OGLE-BLG-RRLYR-05080.} Part of the data in the first season of observations is vertically shifted. We removed these data and analysed the rest. 

\noindent{\bf OGLE-BLG-RRLYR-07653.} After prewhitening the spectrum with a frequency of the first overtone and its harmonics, we still detect residual, unresolved signal at $\nu_{\rm 1O}$. Time dependent analysis shows that amplitude and phase of the first overtone change with similar periodicity. We incorporate the OGLE-III data in the analysis to increase the frequency resolution. Length of the merged OGLE-III and OGLE-IV data is $\approx6042$\thinspace d, which allows to detect modulation period as long as $3020$\thinspace d. Analysis of the combined data shows clear triplets at $\nu_{\rm 1O}$ and its harmonics, which we interpret as Blazhko effect. Period of the Blazhko modulation is $1698 \pm 4$\thinspace d. Additional frequency, at $\approx 0.68\nu_{\rm 1O}$, is visible both in OGLE-IV and in combined data.

\noindent{\bf OGLE-BLG-RRLYR-09146.} 
The additional low frequency signal, $\nu_{\rm x}$, reported in Tab.~\ref{tab:sh} forms two linear combinations with $\nu_{\rm 1O}$ with $S/N>4$ each. Yet another signal ($\nu_{\rm y}$) is visible on the higher frequency side of $\nu_{\rm x}$, at separation $\nu_{\rm y}-\nu_{\rm x}\approx 0.106\,{\rm d}^{-1}$. Period ratio of this third frequency with the first overtone is $0.7245$, too low to consider the additional signal as corresponding to radial fundamental mode. No combinations of $\nu_{\rm 1O}$ and $\nu_{\rm y}$ are visible in the spectrum. 

\noindent{\bf OGLE-BLG-RRLYR-32196.} After prewhitening the data with $\nu_{\rm 1O}$ and its harmonics, additional close peaks (doublets) on the lower frequency side of $\nu_{\rm 1O}$ and $2\nu_{\rm 1O}$ are visible. These peaks may correspond to the Blazhko effect with incomplete triplets (and modulation period of $8.4$\thinspace days) or to a non-radial mode. No OGLE-III data is available for this star.

\section{Summary and conclusions}\label{sec:summary}
We have discovered a new group of double-periodic RR~Lyrae stars with the dominant pulsation in the radial first overtone. Additional period is longer than the first overtone period and longer than the expected period of the unseen fundamental mode. Period ratios between the first overtone and additional period tightly cluster around $0.686$, independently of the first overtone period (which covers a broad range from $\approx0.22$ to $\approx0.42$ d). Amplitude of the additional periodicity is in the mmag range and is only a small fraction of the first overtone amplitude (up to 6 per cent). The group counts 12 stars, 11 identified in the OGLE-IV Galactic bulge photometry and one in {\it Kepler} observations. In two stars we detect the Blazhko modulation. Irregular phase variation, which is typical for RRc stars, is detected in most cases. The signal corresponding to additional periodicity is always stationary.

Excitation of an additional oscillation mode in these RRc stars seems the only possible interpretation of the additional signal. The frequency, though lower than $\nu_{\rm F}$, is still well above the Keplerian frequency and this rules out all interpretations in terms of spots or a companion. We note that instability may extend below the frequency of the fundamental mode, as is the case for some dipole modes plotted in fig.~2 of \cite{dc99}. These modes are of mixed character. To check whether instability extends sufficiently below $\nu_{\rm F}$, so it matches the $\approx 0.68\nu_{\rm 1O}$ modes, a careful analysis of a grid of full evolutionary models is required, which is ongoing (Dziembowski, in prep.). Still, the expected driving rates \citep{vdk98,dc99} are orders of magnitude lower than driving rates for the radial modes.  The driving rates are not a good predictors of the pulsation amplitude or of the form of the finite amplitude pulsation \citep[mode selection, e.g.][]{smolec14}. Still, the orders of magnitude lower driving rates than in apparently not excited fundamental mode are worrying.

 We anticipate a further detections of stars with additional low frequency periodicity during the ongoing K2 mission. The precise space photometry may shed more light on the nature of these mysterious RRc pulsators.

\section*{Acknowledgments}

This research is supported by the Polish National Science Centre through grant DEC-2012/05/B/ST9/03932. Fruitful discussions with Pawel Moskalik are acknowledged. We also thank Jan Skowron for enlightening discussion about the OGLE photometry.

\bsp

\label{lastpage}

\end{document}